\documentstyle[amsfonts,12pt]{article}

\def\D{{\cal D}}
\def\hD{{\hat d}}

\def\Ph{{\hat\phi}}

\def\E{{\rm I}\hskip-.2em{\rm E}}
\def\ra{\rightarrow}
\def\tint{{\textstyle\int}}
\def\hg{{\hat g}}
\def\hP{{\hat p}}
\def\l{\lambda}
\def\hQ{{\hat q}}
\def\hp{{\hat\pi}}
\def\s{\hskip.08em}

\def\a{\alpha}
\def\b{\begin{eqnarray*}}     
\def\e{\end{eqnarray*}}       
\def\bn{\begin{eqnarray}}     
\def\en{\end{eqnarray}}       
\def\<{\langle}
\def\>{\rangle}

\def\{{\lbrace}
\def\}{\rbrace}
\begin{document}
\title{Affine Quantum Gravity\footnote{This essay received 
an ``honorable mention" in the 2003 Essay Competition of the 
Gravity Research Foundation. }}
\author{John R. Klauder
\footnote{Electronic mail: klauder@phys.ufl.edu}\\
Departments of Physics and Mathematics\\
University of Florida\\
Gainesville, FL  32611}
\date{}     
\maketitle
\begin{abstract}
A sketch of the affine quantum gravity program illustrates a 
different perspective on several difficult issues of principle: 
metric positivity; quantum anomalies; and nonrenormalizability.
\end{abstract}
Quantum gravity is under study from several viewpoints. Brane world 
scenarios and loop quantum gravity represent the two most popular 
approaches \cite{bra}, yet alternative viewpoints may also deserve 
consideration. One such alternative --- called affine quantum 
gravity --- rests on just a few basic principles, outlined below. 
First, we recall some hurdles that any approach to quantum gravity must face.

Our list of problems that must be faced is short, but each problem 
is major. These problems are: (1) maintaining the physical nature of 
the metric field; (2) dealing with a set of open first-class classical 
constraints that becomes partially second class on quantization; 
and (3) overcoming the perturbative nonrenormalizability of 
conventional quantum gravity (i.e., Einstein's theory, to which 
we confine our attention).  Affine quantum gravity looks at these 
major problems from a fresh perspective \cite{aff}. What follows is 
a brief overview of this program.

\subsection*{Metric positivity}
Most operator descriptions of quantum theory adopt (or are equivalent 
to) a Hamiltonian point of view. Specifically, there are several 
kinematical operators obeying a basic algebra, and for quantum gravity 
we choose so-called affine commutation
relations, which are not equivalent to canonical commutation relations. 
The determining factor in this choice is ensuring commutation relations 
among self-adjoint field operators that preserve the spectral 
properties implicit in metric positivity (strictly positive spatial 
distances). A one-parameter analog provides a useful clarification. 
Let $\hP$ and $\hQ$ denote Hermitian quantum operators obeying the 
Heisenberg commutation relation $[\hQ,\s\hP]=i\hbar\s I$. Insist that 
$\hQ$ be self adjoint with a positive spectrum, $\hQ>0$. Then $\hP$ 
cannot simultaneously be self adjoint for otherwise it would serve to 
generate unitary translations of $\hQ$ thus violating spectral 
positivity. The remedy is simple: multiply the Heisenberg relation by 
$\hQ$ to obtain $[\hQ,\s\hD]=i\hbar\s\hQ$, where 
$\hD=(\hQ\s\hP+\hP\s\hQ)/2$. Choose the new, affine commutation 
relation as basic; then $\hQ$ and $\hD$ can both be self adjoint 
since $\hD$ serves to dilate $\hQ$, thereby preserving spectral positivity.

Affine commutation relations for the spatial metric tensor operator 
$\hg_{ab}(x)$ $[\s=\hg_{ba}(x)\s]$ and its dilation partner 
$\hp^c_d(x)$ entail self-adjoint field operators (after smearing) 
that have the property of respecting metric positivity. Here 
$\hp^c_d(x)$ is the quantum field associated with the 
classical ``momentric'' field $\pi^c_d(x)=\pi^{cb}(x)\s g_{bd}(x)$, 
where $\pi^{cb}(x)\s [\s =\pi^{bc}(x)\s]$ is the ADM classical 
momentum conjugate to the classical spatial metric tensor 
$g_{ab}(x)\s [\s= g_{ba}(x)\s]$. The affine commutation relations 
for quantum gravity are then nothing more than commutation 
expressions (modulo the usual $i\hbar$) of the associated Poisson 
brackets for the classical fields $g_{ab}(x)$ and $\pi^c_d(y)$ 
that follow directly from the usual Poisson brackets for the classical 
fields $g_{ab}(x)$ and $\pi^{cd}(y)$. Classically, canonical 
kinematical variables ($g_{ab}$ and $\pi^{cd}$) are entirely equivalent 
to affine kinematical variables ($g_{ab}$ and $\pi^c_d$); quantum 
mechanically, only the affine kinematical variables ($\hg_{ab}$ and 
$\hp^c_d$) preserve metric positivity, and their adoption precludes 
the canonical momentum from even exisiting as an operator!

\subsection*{Constraints}
Let us turn attention to a brief account of a unified procedure to  
deal with all kinds of quantum constraints. Classically, constraints 
are a set of real functions, $\{\phi_\a\}_{\a=1}^A$,
of the classical phase space variables, and their vanishing restricts 
the system to a subset of the original phase space, called the constraint 
hypersurface, determined by $\phi_\a=0$ for all $\a$. Quantum 
mechanically, constraints are self-adjoint operators, 
$\{\Ph_\a\}_{\a=1}^A$, that are related to $\{\phi_\a\}_{\a=1}^A$ by
quantization. While $\Sigma_\a\s\phi_\a^2=0$ on the constraint 
hypersurface, it does not follow that $X\equiv\Sigma_\a\s\Ph_\a^2$ 
vanishes on any subspace, but generally $X\ra0$ if $\hbar\ra0$. We 
focus on a subspace of the original Hilbert space, 
$\E\s{\frak H}\subset{\frak H}$, where
$\E=\E(\!\!(\Sigma_\a\s\Ph_\a^2\le\delta(\hbar)^2)\!\!)$ is a projection 
operator that satisfies $0\le \E X\E\le \delta(\hbar)^2I$. By 
choosing $\delta(\hbar)^2$ suitably, various types of constraints 
may be covered, e.g.: (a) if $\Ph_\a=J_\a$, $\a=1,\,2,\,3$, where 
$J_\a$ are generators of SO(3),
then $\delta(\hbar)^2=\hbar^2/2$ forces $J_\a=0$ for all $\a$; (b) 
if $\Ph_1=\hP$ and $\Ph_2=\hQ$, then $\delta(\hbar)^2=\hbar$ leads 
to $\E$ as a projection operator onto states $|\psi\>$ that 
satisfy $(\hQ+i\hP)|\psi\>=0$; (c) if 
$\Ph_1=\hQ$ alone, then we may choose (say) $\delta(\hbar)^2=10^{-1000}$. 
For case: (a) $X$ has a discrete spectrum including zero, and corresponds 
to first-class constraints; (b) $X$ has a discrete spectrum {\it not} 
including zero,
and corresponds to second-class constraints; 
and (c) $X$ has a continuous spectrum including zero, and $\delta$ is 
chosen ridiculously small so that highly unphysical energies would be 
needed to induce excitations above the lowest level for that 
(sub)system (it is also possible to let $\delta\ra0$ as a suitable 
limit to enforce
$\hQ=0$ exactly). These few examples illustrate how general constraint 
operators can be treated within the projection operator formalism 
\cite{op} in a remarkably similar fashion.

For gravity, all physics enters through four constraint fields expressing 
invariance under coordinate transformations. Mutual consistency of the 
constraints as expressed by their Poisson brackets leads to an open set 
of first-class constraints (not characterized by a Lie agebra). When 
quantized, the consistency of these constraints exhibits an anomaly, 
i.e., they are partially second class. While others change the theory 
to avoid this conclusion, we accept it. The projection operator approach 
to quantum constraints outlined above incorporates second-class 
constraints as readily as it does first class ones. Therefore, 
affine quantum gravity formally extends to incorporate the four 
constraint operator fields of gravity. To incorporate such constraints, 
some sort of regularization is required.

\subsection*{Nonrenormalizability}
To deal with nonrenormalizability it helps to understand what it means 
for an interaction to be perturbatively nonrenormalizable. Consider the 
schematic functional integral
\b    S_\l(h)=\int e^{\tint\s h\s f}\;e^{-Q(f)-\l\s N(f)}\;\D f\;,  \e
where $f$ denotes generic fields ($h$ denotes source fields), and $Q$ and
 $N$ denote quadratic and nonquadratic components of the action, 
respectively. If, for all relevant $f$, $Q(f)<\infty\Longrightarrow 
N(f)<\infty$, then $\lim_{\l\ra0^+}\s S_\l(h)=
S_0(h)$, and this situation corresponds to (super)renormalizable 
interactions. However, if $Q(f)<\infty \;\not\!\!\Longrightarrow 
N(f)<\infty$, then
$\lim_{\l\ra0^+}\s S_\l(h)=S'_0(h)\neq S_0(h)$, and this situation 
corresponds to nonrenormalizable interactions. This result holds because 
$N(f)$ acts partially as a hard core in field space, and the 
expression $S'_0(h)$ differs from $S_0(h)$ because it retains the 
hard-core effects even after $\l\ra0^+$. Note that the interacting 
theory is not even continuously connected to the noninteracting theory! 
Consequently, it is not surprising that a power series expansion of a 
(regularized) hard-core, nonrenormalizable interaction leads to an ever 
increasing variety of divergent contributions. 

The hard-core picture for nonrenormalizable interactions is quite 
general \cite{hard}, and formally it applies to various nonrenormalizable 
models, including gravity. However, incorporating a hard core within a 
functional integral
is challenging, and only recently \cite{over} has a fully specific 
proposal been advanced regarding how one might achieve nontrivial results 
for a quartic self-interacting scalar theory in five (or more) spacetime 
dimensions. Demonstration of the hard-core philosophy in such models 
would provide a big boost to their use in other problems such as quantum 
gravity.

\subsection*{Comment}
One of the currently important questions in quantum gravity deals with 
establishing the discreteness of space, and thus the quantization of 
areas and volumes. Affine quantum gravity is not sufficiently advanced 
to address this question, but it is useful to note that there seem to be 
the roots of just such a
topic. Although the metric tensor $\hg_{ab}$ commutes with itself, 
   \b  [\hg_{ab}(x),\s\hg_{cd}(y)]=0 \;, \e
it must be remembered that this metric tensor is not a physical 
observable. Instead,
the physical metric is given by $\hg^E_{ab}(x)\equiv\E\s\hg_{ab}(x)\s\E$,
and it follows almost surely that
   \b [\hg^E_{ab}(x),\s\hg^E_{cd}(y)]\neq 0\;.  \e
While the result of such a commutator is not yet in hand, it is quite
possible that it could lead to discrete units of space.

\subsection*{Summary}
In this brief note we have tried to suggest how one may approach various 
problems in quantum gravity in new ways. First, we have discussed 
the affine commutation relations for the basic kinematical field operators 
that (i) are self-adjoint (after smearing), and (ii) retain the positivity 
of the metric tensor so that spatial distances are strictly positive. 
Second, we illustrated the projection operator approach to quantum 
constraints as a way to accomodate quantum anomalies, i.e., constraints 
that are partially second class, without having to abandon the original 
Einstein theory. Third, we have presented the hard-core theory of 
nonrenormalizability as a way to understand why nonrenormalizable 
interactions behave as they do, as well as to offer, in principle, 
a way to develop a nonperturbative procedure to overcome such problems. 

The program of affine quantum gravity is not easy, but nevertheless 
it offers novel ways to attack some otherwise seemingly intractable problems.

\subsection*{Acknowledgments}
This work has been partially supported by NSF Grant 1614503-12.

\end{document}